
\input phyzzx

\def\mag{$\sp{\vcenter{\hbox{${\scriptstyle
 {\bigtriangleup}}$}\vskip-7.1mm\hbox{${\scriptstyle{\bigtriangledown}}$}}}$}

\def\s{$\sigma$}

\pubnum={19/92}
\date{March 1992}
\titlepage
\title{Quantum Group $\sigma$ Models}
\author{Y. Frishman\mag\ \break
J. Lukierski$\sp{\dagger}$\foot{On leave of absence from the
Institute of Theoretical Physics, University of Wroclaw, ul. Cybulskiego
36, 50-205 Wroclaw, Poland} \break W.J. Zakrzewski $\sp{\ddagger}$     }
\address{\mag\ Department of Physics, Weizmann Institute\break
Rehovot, Israel\break\break
$\sp{\dagger}$Laboratoire de Physique Theorique, Universite de Bordeaux
I,\break
Rue du Solarium 19, F-33175 Gradignan, France\break\break
$\sp{\ddagger)}$Department of Mathematical Sciences\break
      University of Durham, Durham DH1 3LE, England\break\break}
\abstract
Field-theoretic models for fields taking values in quantum groups are
investigated. First we consider $SU_q(2)$ $\sigma$ model ($q$ real)
expressed in terms of basic notions of noncommutative differential
geometry. We discuss the case in which the $\sigma$ models fields are
represented as products of conventional $\sigma$ fields and
of the coordinate-independent algebra. An explicit example is provided by  the
$U_q(2)$ $\sigma$ model with $q\sp{N}=1$, in which case quantum matrices
$U_q(2)$ are realised as $2N\times 2N$ unitary matrices. Open problems
are pointed out.

\endpage
\chapter{Introduction}
\REF\rTak{L. Takhtajan, L.D. Faddeev - {\it Russian Math. Surveys} {\bf
34}, 11 (1979)}
\REF\rSkyanin{E.K. Sklyanin and P.P. Kulish - {\it Lecture Notes in
Physics} {\bf 151}, 61 (1982)}
\REF\rFad{L.D. Faddeev - {\it Les Houches Lectures 1982}, ed. Elsevier,
Amsterdam (1984)}\REF\rFAD{L.D. Faddeev, N.Yu. Reshetikin and L.A. Takhtajan -
 {\it Algebra and Analysis} {\bf 1}, 178 (1989)}\REF\rW{S.L. Woronowicz -
{\it Publ. RIMS, Kyoto Univ.} {\bf 23}, 117 (1987)}\REF\rWor{S.L.
Woronowicz - {\it Comm. Math. Phys}
 {\bf 111}, 613 (1987)}
 The appearance of noncommutative entries in some matrices describing
 quantum inverse scattering methods for spin systems (see \eg\
 [\rTak-\rFad]) has led to the introduction of the concept of a quantum
 matrix group.\refmark\rFAD$\sp{-}$\refmark\rWor\   From the
algebraic point of view the description of a quantum group as a
quasitriangular Hopf algebra was first given by Drinfeld\Ref\rD{V.G.
Drinfeld - {\it Proc of XX International Math. Congress, Berkeley} Vol.I
, 798 (1986)}\rlap, with the basic object being the noncommutative algebra of
functions on a quantum group. The quantum extensions of all classical
matrix groups (Cartan $A_n$, $B_n$, $C_n$ and $D_n$ series), describing the
generators of Drinfeld's quantum algebra have also been
given\refmark\rFAD\rlap.
\REF\rPod{P. Podles - {\it Lett. Math. Phys.} {\bf 14},
193 (1987); ibid, {\bf 18},
107 (1980)}\REF\rNou{M. Noumi and K. Mimachi - {\it Comm. Math. Phys.} {\bf
128}, 521 (1990)}\REF\rYam{M. Noumi, H. Yamada and K. Mimachi - {\it Proc.
Jap. Acad. Sci.} {\bf 65A}, 169 (1989)}

Moreover, the quantum counterparts of the homogeneous coset spaces
(\eg\ spheres - $S_q\sp{n}$, projective spaces - $CP_q(n)$ \etc) have also been
found\refmark\rFAD\refmark\rPod$\sp{-}$\refmark\rYam.
 \REF\rz{W.J. Zakrzewski - {\it Low dimensional $\sigma$ models} - Adam
Hilger (1989)}

\par In this paper we will consider fields taking values in quantum groups
and we will discuss the corresponding $\sigma$ models. Let us recall that
the usual $\sigma$-field $\phi(x)$ describes the mappings from the
coordinate manifold $S$ into the target space $M$ (see \eg\ [\rz]). In
principle we can ``$q$-deform" the target space $M$ ($M\rightarrow M_q$)
as well as the coordinate manifold $S$ ($S\rightarrow S_q$), \ie\ we can
introduce three kinds of $\sigma$ models: \break
\vskip-8mm\item{a)} with quantum deformation of the target manifold
$$\phi\sp{q}_a(x):\qquad x\in S\;\rightarrow\;\phi_a\sp{q}\in M_q,
\eqn\eone$$ where the index $a$  ennumerates the local coordinates
on $M_q$.
\item{b)} with quantum deformation of the base manifold
$$\phi_a(x_q):\qquad x_q\in S_q\;\rightarrow\;\phi_a\in M,\eqn\oneb$$
\item{c)} with both manifolds deformed
$$ \phi\sp{q}_a(x_q):\qquad x_q\in S_q\;\rightarrow\;\phi_a\sp{q}\in M_q.
\eqn\eonec$$

Let us add that, analogously, there exist three types of supersymmetric
models, corresponding to the three types of mappings, and that we believe
that in the near future, the quantum group $\sigma$ models of all three
types will be studied. In the present paper we will discuss mainly the case
\eone.
\par The description of quantum group $\sigma$ models can be presented in
two different ways: \hfil\break
1) We can consider quantum group $\sigma$ fields as fields satisfying
at a point $x\in S$ the quantum algebra, and so we can study the general
properties which follow from the basic formulae of noncommutative
differential geometry on quantum groups. Such an approach, which we shall
call algebraic, was recently used by Arefeva and Volovich\Ref\rA{I.Y.
Arefeva and I.V. Volovich - {\it Mod. Phys. Lett.} {\bf 6A}, 893
(1991)}\Ref\rV{I.Y. Arefeva and I.V. Volovich - {\it Phys. Lett.} {\bf
264B}, 62 (1991)}.\REF\rWW{S.L. Woronowicz - {\it Comm. Math. Phys.}
 {\bf 122}, 125 (1989)}
 In the algebraic formulation of a quantum group $\sigma$
model one can repeat the major part of the geometric formulation of the
standard approach to the $\sigma$ models (Cartan forms, Cartan structure
relations, algebra of the covariant derivatives \etc), provided that the
exchange relations between quantum group valued $\sigma$ fields and their
derivatives are properly introduced. For the $SU_q(2)$ case the formulae of
noncommutative geometry are well known (see \eg\ ref [\rW,\rWW]) and so in
the next section we shall present the algebraic formulation of the $SU_q(2)$
$\sigma$ model. \hfil\break
2) We can assume that the quantum group $G_q$ $\sigma$ fields
$\phi_A\sp{q}(x)$ are products of ``ordinary" functions $f_A(x)\in H$ and
of the $x$ independent algebra $A$ related to the quantum group algebra.
This approach gives us expressions which belong to the tensor product
$H_G\otimes T_A$, where for an $n\times n$ matrix quantum group $\sigma$
model the first part $H_G$ is parametrised by an $n\times n$ matrix of
``classical" fields (suitably constrained standard $GL(n)$ $\sigma$ fields),
and  $T_A$ carries the realisation of the algebra $A$. If $A=f(G_q)$, the
natural realisation on the polynomial basis of the functions on the quantum
group is infinite dimensional, and for $q$ real it can not be reduced to a
finite dimensional case. In the second part of section 2 we will present
our discussion of the $SU_q(2)$ quantum group $\sigma$ model for the
solutions satisfying the separability condition described above.
\REF\rc{C. de Concun and V.G. Kac - ''Representations of quantum groups at
root of 1", {\it Progress in Math.} Vol {\bf 92}, 471 (1990); ed.
Birkhauser (Basel)}
\REF\rDa{E. Date, M. Jimbo, K. Miki and T. Miwa - {\it Com. Math. Phys.}
{\bf 137}, 133 (1991)}
\REF\rAa{D. Arnandon and A. Chakrabarti - {\it Com. Math. Phys.} {\bf 139},
605 (1991)}
\par The realisation $T_A$ can be described by finite matrices in one case:
when $q$ is complex and $q\sp{N}=1$ ($q=e\sp{i{2\pi\over N}}$). It is worth
mentioning that some realisations of the quantum groups for $q$ being the
$N$-th root of unity have recently been found to have physically relevant
applications (see \eg\ ref [\rc,\rDa,\rAa]). It appears, however, that the
quantum deformations $O_q(n)$ and $SU_q(n)$ of the semisimple groups $O(n)$ and
$SU(n)$  which are the natural candidates for quantum group $\sigma$ fields,
 do not
permit complex $q$\refmark\rFAD.  \par The simplest example of a compact
quantum group with $q\sp{N}=1$, the  quantum group $U_q(2)$, will be considered
in section 3. There, we will first show that the $U_q(2)$ for complex
values of $q$ can be obtained as a special case of a two-parameter
deformation $GL_{q,p}(2,C)$ of the $2\times 2$ general linear group. Then,
using the results on the matrix realisations of $GL_q(n)$ for
$q\sp{N}=1$\Ref\rF{E.G. Floratos - {\it Phys. Lett.} {\bf B233}, 395
(1989)} \Ref\rWey{J. Weyers - {\it Phys. Lett.} {\bf B240}, 396( 1990)}
we will embed the $U_q(2)$ $\sigma$ model with $q\sp{n}=1$ (``anyonic
$\sigma$ model") into the conventional $U(2n)$ $\sigma$ model. It appears
that when we use such a representation we describe solutions that,
 for $D=1$ (see eq. (3.2)), satisfy
the assumptions made by Arefeva and Volovich in
their discussion of quantum group sigma models\refmark\rA. Finally, some
open problems are discussed in section 4.
\REF\rVaks{L.L. Vaksman and Y.S. Sorbelman - {\it Funct. Anal. and its
Appl.} {\bf 22}, 170 (1989)}
\REF\rWess{A. Schirrmecher, J. Wess and B. Zumino - {\it Zeit. f. Phys.}
{\bf C49}, 317 (1991)}
\REF\rEguchi{T. Eguchi and H Kawai, {\it Phys. Lett.} {\bf B110}, 143 (1982)}
\REF\rKoorn{T.H. Koornwinder - {\it Neder. Acad. Wetensch. Proc. Ser.} {\bf
A92}, 97 (1989)}
\REF\rBaulieu{L. Baulieu and E.G. Floratos - {\it Phys. Lett.} {\bf B258},
 171-178 (1991)}
\REF\rTze{F. Gursey and H.C. Tze - {\it Ann. of Phys.} {\bf 128}, 29
(1980)}
\REF\rreview{see \eg\ C.M. Hull, {\it Lectures on nonlinear $\sigma$ models},
in {\it Super Field Theories}, New York, Plenum (1987)}

\chapter{$SU_q(2)$\ \s\ model}
\section{Algebraic formulation}
Let us first introduce the quantum group $SL_q(2,C)$ as the following Hopf
bialgebra ($q$ complex):\refmark\rFAD \hfil\break
a) multiplication:
$$\eqalign{U\,&=\,\pmatrix{a&b\cr c&d\cr};\cr
ab\,&=\,qba\qquad ac\,=\,qca\qquad cd\,=\,qdc\cr
bc\,&=\,cb \qquad bd\,=\,qdb\cr
ad\,&-\,qbc\,=\,da-q\sp{-1}cb\,=\,1.\cr}\eqn\etwoone$$
b) coproduct
$$\Delta\pmatrix{a&b\cr c&d\cr}\,=\,\pmatrix{a&b\cr c&d\cr}\times \pmatrix
{a&b\cr c&d\cr},\eqn\etwotwo$$
c) inverse (antipode) and co-unit
$$S\pmatrix{a&b\cr c&d\cr}\,=\,\pmatrix{d&-q\sp{-1}b\cr -qc&a\cr}\qquad
\epsilon\pmatrix{a&b\cr c&d\cr}\,=\,\pmatrix{1&0\cr 0&1\cr}.\eqn\ethree$$
For $q$ real we can introduce the following unitarity condition. For
$$U\,=\,\pmatrix{u_{11}&u_{12}\cr u_{21}&u_{22}\cr}=\pmatrix{a&b\cr
c&d\cr}\eqn\eaaaa$$ we put
$$U\sp{\dagger}\;=\;\pmatrix{a\sp{\star}&b\sp{\star}\cr
c\sp{\star}&d\sp{\star}\cr}\sp{T}\;=\;\pmatrix{d&-q\sp{-1}b\cr
-qc&a\cr}\;=\;S(U),\eqn\etwofour$$
which defines the $SU_q(2)$ quantum group, the matrix elements of which
have the form $$U=\pmatrix{a&-qc\sp{\star}\cr c&a\sp{\star}\cr}.\eqn\eee$$
In this case the relations \etwoone\ take the form
$$\eqalign{ac\,=&\,qca\qquad ac\sp{\star}\,=\,qc\sp{\star}a\qquad
cc\sp{\star}\,=\,c\sp{\star}c\cr aa\sp{\star}+&q\sp2 cc\sp{\star}=1\qquad
a\sp{\star}a+c\sp{\star}c=1.\cr}\eqn\etwofive$$
\par In order to define a quantum group \s\  model we introduce the Cartan
one-forms on $SU_q(2)$
$$\Omega\,=\,U\sp{\dagger}dU\,\leftrightarrow\,\Omega_{ik}\,=\,U_{ij}
\,dU_{jk}.\eqn\esix$$
The formula \esix\ describes the left-invariant one-forms
($\Omega=\Omega_l$). The right-invariant forms are given in terms
of the left-invariant forms in the same way as in the $q=1$
case, and so are given by
$$\Omega_R\;=\;-U\Omega_lU\sp{\dagger}.\eqn\eright$$
We  also have
$$\eqalign{dU\,&=\,U\Omega_l\,=\,-\Omega_RU,\cr
dU\sp{\dagger}\,&=\,U\sp{\dagger}\Omega_R\,=\,-\Omega_lU\sp{\dagger}.\cr}
\eqn\esixc$$
For the Cartan one-form \esix\ we can introduce the linear basis.
Following the so-called $4D_+$ bi-covariant calculus of
 Woronowicz\refmark\rPod\refmark\rWW\  we can choose
$\Omega=\omega_A\tau_A$ ($A=0,1,2,3)$, where
$$\tau_1\,=\,\pmatrix{0&1\cr 0&0\cr}\qquad \tau_2\,=\,\pmatrix{0&0\cr
{-1\over q}&0\cr}\quad \tau_3\,=\,{1\over 1+q\sp2}\pmatrix{-{1\over q}&0\cr
0&{1\over q}\cr}\eqn\etwoseven$$
and $\tau_0={q\sp3-1\over (1+q)(1+q\sp2)}I,$ where $I$ denotes the unit matrix.
Observe that we have chosen a nonsingular normalisation (compare with ref
 [\rPod], second paper, p.108).

\par The $SU_q(2)$ \s\ fields are then introduced by the mapping (1.1)
\ie\ $U_{ij}\rightarrow U_{ij}(x)$. If we now write
$$ \eqalign{\Omega_{ik}&=U_{ij}\sp{\dagger}{\partial U_{jk}\over \partial
 x_{\mu}
} dx_{\mu}\,=\,\Omega_{ik}\sp{\mu}dx_{\mu},\cr \Omega_{\mu}\,=&\,\omega_{A\mu}
\tau_{A},\cr}
\eqn\etwoeight$$
the action of the $SU_q(2)$ model can be written as:
$$\tilde S\,=\,-\int d\sp{d}x\,\Tr{(\Omega_{\mu}\Omega\sp{\mu})}\,=\,-
\int\,d\sp{d}x\,G_q\sp{AB}\omega_{A\mu}\omega_B\sp{\mu},\eqn\etwonine$$
where $G_q\sp{AB}=\Tr{(\tau_A\tau_B)}$ is given by
$$G_q\sp{AB}\;=\;\pmatrix{0&-{1\over q}&0&0\cr
-{1\over q}&0&0&0\cr 0&0&{2\over q\sp2(1+q\sp2)\sp2}&0\cr
0&0&0&{2(q\sp3-1)\sp2\over (1+q)\sp2(1+q\sp2)\sp2}\cr}.\eqn\etwoten$$
The Cartan forms $\omega_A$ describe the \s\ model currents.
We see that the contribution of the scalar current $\omega_0$ vanishes in
the
limit $q\rightarrow 1$. $G_q\sp{00}$ vanishes
at $q=1$ by the choice of $\tau_0$. This choice is consistent as
 in this limit we get the $SU(2)$ model and no
 $U(1)$ current.
\par We can now consider \s\ fields and take currents $\omega_A$ as our
basic variables.
Due to the unitarity ($\Omega=U\sp{\dagger}dU=-dU\sp{\dagger}U$) we can
rewrite the action \etwonine\ as
$$\tilde S=\int d\sp{d}x{\partial U_{ji}\sp{\star}\over \partial
x\sp{\mu}}{\partial
U_{ji}\over \partial x_{\mu}}\,=\,\int d\sp{d}x{\partial(SU)_{ij}\over\partial
x\sp{\mu}}{\partial U_{ji}\over \partial x_{\mu}}.\eqn\etwoeleven$$
Denoting $$U(x)=\pmatrix{A(x)&B(x)\cr C(x)&D(x)\cr}\eqn\ebbb$$ we obtain
$$\tilde S\,=\,\int
d\sp{d}x(A_{,\mu}\sp{\star}A\sp{,\mu}+q\sp2C_{,\mu}C\sp{\star,\mu}
+C_{,\mu}\sp{\star}C\sp{,\mu}+A_{,\mu}A\sp{\star,\mu},\eqn\etwotwelve$$
where the field operators $A(x),A\sp{\star}(x),C(x),C\sp{\star}(x)$
 satisfy the algebra
\etwofive\ at every coordinate point $x$.  But as we have the operators
and their derivatives, we  need to know the algebra
at points $x$ and $x+\epsilon$,  with $\epsilon$ infinitesimal.
To do this we have
to
determine
the complete algebra for our basic fields, \ie\ for $U_{ij}$ and $\omega_A$.
This algebra, in the case of $SU_q(2)$, is known in an explicit
form\refmark\rW\refmark\rWW.
\hfil\break

\section{Separable realisations}
Let us assume that
$$U_{ij}(x)\,=\,f_{ij}(x)\otimes \hat U_{ij},\quad
i,j=1,2\eqn\etwotheirteen$$
where the functions $f_{ij}(x)$ are classical and $\hat U_{ij}$ describe the
coordinate independent operators. Further, let us assume that the quantum
\s\ field \etwotheirteen\ satisfies the unitarity condition $UU\sp{\dagger}=
U\sp{\dagger}U=1.$
\par We shall consider here only two cases:\hfil\break
$\alpha$) The operators $\hat U_{ij}$ describe the generators of the
$SU_q(2)$ quantum algebra \etwofive\ separately (obviously the total $U_{ij}$
should).
\par In this case the unitarity condition, with  $f_{11}=f$, $f_{12}=g$,
$f_{21}=h$, $f_{22}=k$ and $ \hat c=\hat u_{21}$ is
$$ \eqalign{h\,=\,g\sp{\star},\quad& k\,=\,f\sp{\star}\cr
\vert f\vert \sp2(1-q\sp2\hat c\sp{\dagger}\hat c)\,&+\,q\sp2\vert g\vert\sp2
\hat c\sp{\dagger} \hat c\,=\,1\cr
\vert f\vert \sp2(1-\hat c\sp{\dagger}\hat c)\,&+\,\vert g\vert\sp2
\hat c\sp{\dagger}\hat c\,=\,1.}\eqn\etwofourteen$$
We see that for $q\ne1$ we have
$$\vert f\vert \sp2\,=\,\vert g\vert \sp2\,=\,1, \eqn\etwofifteen$$
\ie\ we obtain the $U(1)\times U(1)$ classical \s\ model.\hfil\break
$\beta$) The functions $f_{ij}(x)$ describe the $SU(2)$ \s\ fields and
so besides the conditions in the first line of \etwofourteen\
they satisfy also $$ \vert f\vert \sp2\,+\,\vert g\vert
\sp2\,=\,1.\eqn\etwosixteen$$
\par Then, we can write \etwotheirteen\ as an $SU_q(2)$ matrix
$$U\;=\;\pmatrix{f(x)\cdot A&-qg(x)\cdot C\sp{\dagger}\cr
g\sp{\star}\cdot C&f\sp{\star}\cdot A\sp{\dagger}\cr}, \eqn\twoseventeen$$
where $A$ and $C$ do not depend on $x$. Thus
we obtain for $A,$ $A\sp{\dagger}$, $C$ and $C\sp{\dagger}$ the relations
of the first line of \etwofive\  and
$$\eqalign{\vert f\vert \sp2 AA\sp{\dagger}\,+&\,q\sp2\vert g\vert \sp2
CC\sp{\dagger}\,=\,1\cr
\vert f\vert \sp2 A\sp{\dagger}A\,+&\,\vert g\vert\sp2
C\sp{\dagger}C\,=\,1,\cr}\eqn\twoeightteen$$
 whose solutions exist only for $q=1$, and only when
$AA\sp{\dagger}=A\sp{\dagger}A=CC\sp{\dagger}=1$.
\par Thus, in conclusion, we see that the assumption \etwotheirteen\ for
$q$ real and with either case $\alpha)$ or $\beta)$,
 is too restrictive and should be extended to, say
$$U_{ij}(x)\;=\;\sum_n\,f_{ij}\sp{(n)}(x)\otimes \hat
U_{ij}\sp{(n)},\eqn\etwonineteen$$
where $\hat U_{ij}\sp{(n)}$ describe a polynomial basis of the ring
of noncommutative functions $f(G_q)$ (see \eg\ ref [\rVaks]).
Such an assumption corresponds to the considering of the mapping
(1.2) with $S_q=S\otimes G_q$, and $M=GL(2)$. \hfil\break
\section{Embeddings in the $U(\infty)$ \s\ model}
Another way of representing  the operators $A$,
$A\sp{\dagger}$, $C$ and $C\sp{\dagger}$ of \etwotwelve\ corresponds to the
use of the parameter dependent irreducible representations of the
functions $f(SU_q(2))$ in a separable Hilbert space $H$. Promoting the
parameters to the functions generates an $\infty$-dimensional \s\ model.
\par The irreducible representations of the $SU_q(2)$ algebra in a Hilbert
space are known\refmark\rW\refmark\rVaks. There are only 2 series of
irreducible representations of $f(SU_q(2))$, each one parametrised
by the parameter of the unit circle $t=e\sp{i\phi}\in S\sp2$. One series
is degenerate, as for it  only the element $a$ of \etwofive\ is
represented in a nontrivial way. The other irreducible representation
is nontrivial. It is described by the operators $\rho_{\phi}$
which act as
$$\eqalign{
\rho_{\phi}(a)e_0=0\quad&\rho_{\phi}(a)e_k=(1-q\sp{-2k})\sp{1\over2}
e_{k-1}\cr
\rho_{\phi}(c)e_k\,=&\,(e\sp{i\phi}\cdot q\sp{-k})e_k,\cr}\eqn\etwotwenty$$
where $e_k$, for  $k=0,1,\cdots\infty$ describes an orthonormal basis in $H$.
\par If we now replace $\phi$ by a function $\phi(x)$ we obtain from
\etwotwenty\ the embedding of the $U(1)$ \s\ field into $U(\infty)$, in
analogy with the separable realisations \etwotheirteen.

\chapter{$U_q(2)$ \s\ model for $q\sp{N}=1$}
In order to obtain a solution of a quantum group \s\ model in the separable
form (see \etwotheirteen), it will
turn out that  we should consider the deformation parameter
$q$ as being complex and satisfying $\vert q\vert=1$. Then, if $q$ equals
the $N$-th root of unity ($q\sp{N}=1$) one can represent the operators
$\hat U_{ij}$ by $N\times N$ dimensional matrices (see refs [\rF], [\rWey]).
As we want to consider a quantum group \s\ model defined on a quantum
compact group, we shall discuss here the simplest such compact quantum group
permitting complex values of $q$, namely, the quantum group $U_q(2)$.
\section{Quantum group $U_q(2)$}
We shall define the quantum group $U_q(2)$, for complex $q$,
 as a real form of a two parameter
deformation of $GL(2,C)$, denoted by $GL_{p,q}(2)$\refmark\rWess.
Then we will know that the real Hopf algebra is valid for our $U_q(2)$.
\par The formulae \etwoone\ have to be extended and they become
$$\eqalign{ab\,&=\,pba\quad ac\,=\,qca\quad cd\,=\,pdc\cr
pbc\,&=\,qbc\qquad bd\,=\,qdb\cr
ad\,-&\,da\,=\,(p-q\sp{-1})bc\cr}\eqn\ethreeone$$
with a coproduct still defined by \etwotwo. If we now introduce
the determinant
$$\eqalign{D\,=&\,ad-pbc\,=\,ad-qcb\cr
=&\,da-p\sp{-1}cb\,=\,da-q\sp{-1}bc\cr}\eqn\ethreetwo$$
then one can check from \ethreeone\ that the following relations hold:
$$\eqalign{[D,\,a]=0,&\qquad [D,\,d]=0\cr
qDb=pbD,&\qquad pDc=qcD.\cr}\eqn\threethree$$
We see that only if $q=p$ we can put $D=1$, \ie\ we have the quantum
group $SL_q(2)$ defined by \etwoone.
\par The quantum group $GL_{p,q}(2)$ is a genuine Hopf algebra for any
complex $q$ and $p$. In particular, the antipode of $$U=\pmatrix{a&b\cr
c&d\cr}\eqn\ezzz$$ is given by the formulae
$$S\pmatrix{a&b\cr c&d\cr}=D\sp{-1}\pmatrix{d&-q\sp{-1}b\cr
-qc&a\cr}=\pmatrix{d&-p\sp{-1}b\cr -pc&a\cr}D\sp{-1},\eqn\ethreefour$$
where we have used $DD\sp{-1}=D\sp{-1}D=1$.
If we now impose the unitarity condition $U\sp{\dagger}=S(U)$, \ie\
$$\eqalign{a\sp{\star}\,&=\,D\sp{-1}d\,=\,dD\sp{-1}\cr
b\sp{\star}\,&=\,-qD\sp{-1}c\,=\,-pcD\sp{-1}\cr
c\sp{\star}\,&=-q\sp{-1}D\sp{-1}b\,=\,-p\sp{-1}bD\sp{-1}\cr
d\sp{\star}\,&=\,D\sp{-1}a\,=\,aD\sp{-1},\cr}\eqn\ethreefive$$
we find as the consistency conditions, that
$$D\sp{\star}\,=\,D\sp{-1},\qquad p=q\sp{\star}\eqn\ethreesix$$
and so  obtain the following $U_q(2)$ algebra
$$\eqalign{ac=qca,&\qquad ac\sp{\star}=q\sp{\star}c\sp{\star}a,\cr
&c\sp{\star}c=cc\sp{\star}.\cr}\eqn\ethreeseven$$
Moreover, from \ethreetwo\ and \ethreefive\ we find that
$$a\sp{\star}a+c\sp{\star}c=1,\qquad aa\sp{\star}+\vert
q\vert\sp2cc\sp{\star}=1.\eqn\ethreeeight$$
\par
\noindent  The quantum matrix $U_q(2)$  given by
$$U\;=\;\pmatrix{a&-q\sp{\star}c\sp{\star}D\cr c&a\sp{\star}D\cr}.
\eqn\ethreenine$$
 It describes the generators of the Hopf algebra with standard comultiplication
rule
$$\Delta(U_{ik})\,=\,\sum_{j=1,2}U_{ij}\otimes U{jk}\eqn\ethreeten$$
and the following antipode condition (using $qD\sp{-1}c=q\sp{\star}cD\sp{-1}$)
$$U\sp{\dagger}\,=\,S(U)\,=\,\pmatrix{a\sp{\star}&c\sp{\star}\cr
-qD\sp{-1}c&D\sp{-1}a\cr}.\eqn\ethreeeleven$$
It should be stressed that even for $\vert q\vert=1$ but with $q\ne1$, we
cannot
put $D=1$ as in this case \ethreeeight\ would give
$$a\sp{\star}a+c\sp{\star}c=1 \qquad
a\sp{\star}a=aa\sp{\star}.\eqn\ethreetvelve$$

\section{$2N\times 2N$ matrix realization of $U_q(2)$ for $
q\sp{N}=1$}
If  $ q\sp{N}=1$ the elements $a$, $c$, $a\sp{\star}$ and
$c\sp{\star}$ can be represented by $N\times N$ matrices.
Following ref. [\rF,\rWey] we introduce the following matrices\foot{The
matrices (3.13) were introduced earlier by Eguchi and Kawai in their
construction of the ''twisted Eguchi-Kawai" models\refmark\rEguchi.}
$$Q\,=\,\pmatrix{q&..&0\cr 0&\ddots&0\cr 0&..&q\sp{N}\cr}\qquad
P=\pmatrix{0&1&..&0\cr
0&0&\ddots&0\cr 0&0&..&1\cr 1&0&..&0\cr}.\eqn\ethreethirteen$$
These matrices satisfy $PQ=qQP$, and with $q=exp({2\pi i\over N})$
they generate the algebra of all $N\times N$ matrices.
Moreover, as $Q\sp{N}=P\sp{N}=1$ we find that $Q\sp{\dagger}=Q\sp{N-1}$
and $P\sp{\dagger}=P\sp{N-1}$ and, as is easy to see,
$$q\sp{ij}Q\sp{i}P\sp{j}\,=\,P\sp{j}Q\sp{i}.\eqn\ethreefourteen$$
This observation suggests that if we restrict ourselves to the $q$'s being
the $N$'th root of identity, we can seek a solution of our conditions
\ethreeseven-\ethreeeight\ with the elements of $U$ given in terms of
various linear combinations of the products of $Q$'s and $P$'s. The
simplest of such
solutions corresponds to the case when $a\sim Q$ and $c\sim P$ and so is
given by
$$\eqalign{a=sin\alpha e\sp{i\psi}P,&\quad a\sp{\star}=sin\alpha
e\sp{-i\psi} P\sp{N-1}\cr
c=cos\alpha e\sp{i\phi}Q,&\quad c\sp{\star}=cos\alpha
e\sp{-i\phi}Q\sp{N-1},\cr}\eqn\ethreefifteen$$
where $\alpha$ and $\psi$ are real numbers. Then it is easy to check that
all conditions (3.7) and (3.8) (with $\vert q\vert=1$) are
satisfied if we choose $$D\,=\,e\sp{i\xi}P\sp2.\eqn\ethreefifteenb$$
Hence a $2N\times 2N$ matrix \ethreenine\ with its entries given by
\ethreefifteen\ and \ethreefifteenb\ is a representation of $U_q(2)$
for $q=exp({2\pi i\over N})$.
\par In the following we shall express the matrix \ethreenine\ as a product
of two matrices $U=T\tilde D$, $U\sp{\dagger}=\tilde
D\sp{\dagger}T\sp{\dagger}$, where
$$T\,=\,\pmatrix{sin\alpha e\sp{i\psi}P& -q\sp{\star}cos\alpha e\sp{-i\phi}
Q\sp{\dagger}\cr cos\alpha e\sp{i\phi}Q& sin\alpha
e\sp{-i\psi}P\sp{\dagger}\cr}\eqn\ethreesizteena$$
and $$\tilde D\,=\,\pmatrix{1&0\cr
0&e\sp{i\xi}P\sp2\cr}.\eqn\ethreesixteenb$$
 The basic Lagrangian then becomes
$$L=-{1\over
2}\Tr{(U\sp{\dagger}\partial_{\mu}U)(U\sp{\dagger}\partial\sp{\mu}U)}\eqn
\ethreeseventeen$$ with the trace taken with respect to the $U_q(2)$
matrix indices as well as the ones describing the realisations
(3.18) and (3.19). One can write
$$\Omega\sp{\mu}\,=\,U\sp{\dagger}\partial\sp{\mu}U\,=\,\tilde D\sp{\dagger}L
\sp{\mu}\tilde D\,+\,\tilde D\sp{\dagger}\partial\sp{\mu}\tilde D,
\eqn\ethreeeighteen$$
where
$$L\sp{\mu}\,=\,T\sp{\dagger}\partial\sp{\mu}T\,=\,\pmatrix{V\sp{\mu}&
Z\sp{\mu}\cr K\sp{\mu}& -V\sp{\mu}\cr}.\eqn\ethreenineteen$$
We obtain
$$V\sp{\mu}\,=\,i(\psi,\sp{\mu}sin\sp2\alpha\,+\,\phi,\sp{\mu}cos\sp2
\alpha)\eqn\ethreetwenta$$
and
$$\eqalign{Z\sp{\mu}\,=&\,Q\sp{\dagger}P\sp{\dagger}y\sp{\dagger \mu}
=Q\sp{\dagger}P\sp{\dagger}e\sp{-i\psi-i\phi}[\alpha,\sp{\mu}
-i(sin\alpha cos\alpha)(\psi,\sp{\mu}-\phi,\sp{\mu})]\cr
K\sp{\mu}\,=&\,-PQy\sp{\mu}=-PQe\sp{i\psi+i\phi}[\alpha,\sp{\mu}
+ i(sin\alpha
cos\alpha)(\psi,\sp{\mu}-\phi,\sp{\mu})]=-Z\sp{\mu\dagger}
\cr}\eqn\ethreetwentyb$$
where
$$\rho,\sp{\mu}={\partial \rho\over \partial
x\sp{\mu}}.\eqn\ethreetwentyone$$
Hence $L\sp{\mu}$ can be resolved into
$$L\sp{\mu}=V\sp{\mu}\pmatrix{1&0\cr0&-1\cr}+y\sp{\dagger\mu}
\pmatrix{0&Q\sp{\dagger}P\sp{\dagger}\cr0&0\cr}-y\sp{\mu}\pmatrix{0&0\cr
PQ&0\cr}\eqn\ethreetwentyonea$$
and so we observe the explicit appearance of the $SU_q(2)$ algebra
generators (which in this case are also the $SU(2)$ generators).
Notice that as $D=e\sp{i\xi}P\sp2$
$$\Omega\sp{\mu}\,=\,\pmatrix{V\sp{\mu}&Z\sp{\mu}D\cr D\sp{-1}K\sp{\mu}&
-\tilde V\sp{\mu}\cr},\eqn\ethreetwentytwo$$
where $$\tilde
V\sp{\mu}=V\sp{\mu}-D\sp{-1}\partial\sp{\mu}D
=V\sp{\mu}-i\partial\sp{\mu}\xi\eqn\ethreetwentythree$$
and the Lagrangian (3.20) is given by the formula
$$L=-{1\over 2}V\sp{\mu}V_{\mu}+y\sp{\mu}y\sp{\dagger}_{\mu}
+{1\over 2}\xi_{,\mu}\xi,\sp{\mu}-iV\sp{\mu}\xi_{,\mu}\eqn\threetwentufour$$
where $\xi(x)$ describes the $U(1)$ field extending $SU(2)$ to $U(2)$.
Indeed, one can show that if we put $\xi=0$ we recover the classical
action for the $SU(2)$ \s\ model. On the other hand, we can generalise
the Lagrangian \ethreeseventeen\ to
$$L\,=\,-{1\over 2}\Tr{(U\sp{\dagger}}\bigtriangledown_{\mu}U)(U\sp{\dagger}
\bigtriangledown\sp{\mu}U)\,=\,-\Tr{}{1\over
2}\tilde\Omega_{\mu}\tilde\Omega\sp{\mu},\eqn\ethreetwentyfive$$
where we have introduced the $U(1)$ covariant derivative
$$\bigtriangledown_{\mu}\,=\,\pmatrix{\partial_{\mu}\cdot 1_N&0\cr
0&(\partial_{\mu}-A_{\mu})\cdot 1_N\cr},\eqn\ethreetwentysix$$
and so
$$\tilde\Omega\sp{\mu}\,=
\pmatrix{{V\sp\mu}&{Z\sp\mu
 D}\cr{D\sp{-1}K\sp\mu}&
 {-(\tilde{V}\sp\mu+A\sp\mu)}\cr}.\eqn\ethreetwentyseven$$

Then, if in particular, we choose the pure gauge mode for the $A_{\mu}$
field
$$A_{\mu}\,=\,D\sp{-1}\partial_{\mu}D\,=\,i\xi_{,\mu}\cdot
1_N,\eqn\ethreetwentyeight$$
we find that \ethreetwentyfive\ reduces to the conventional $SU(2)$
\s\ model. We see that the $U(1)$ gauge field $A_{\mu}$ leads to the
appearance of the gauge invariance which allows us to set $D=1$ in
\ethreetwentytwo.

\chapter{Outlook}
In this paper we have considered some particular solutions of $\sigma$ models
taking values in quantum groups. Our main example corresponded to
the $U_q(2)$ \s\ model with $q\sp{N}=1$. The ``classical" fields of this model
were described by $2N\times 2N$ matrices, \ie\ we considered the embedding
 $U_q(2)\subset\,
   U(2N)$. In the general case, however, the embeddings must involve infinite-
dimensional \s\ models \ie\ $U(\infty)$ or $O(\infty)$. Indeed, the quantum
 group generators can be represented in terms of the Heisenberg algebra, which
 can be
realised using infinite-dimensional matrices. In  fact, it is only when
we  impose the relation
$$[a,\,a\sp{\star}]=0\eqn\eCOM$$
that we can realise the generators of $SU_q(2)$ or $U_q(2)$ in terms of finite
dimensional matrices. For $SU_q(2)$, eqs. (2.7) and (4.1)
imply $q\sp2=1$. For $U_q(2)$, eqs. (3.9) and (4.1) imply $\vert q\vert=1$.
\par We hope that our paper will be treated as a preliminary study of some
 aspects of sigma models taking values in a quantum group.
Let us mention some of these aspects:\hfil\break
a) For a given quantum group one can consider any finite dimensional
 representation ($u_{ij}\rightarrow T_{AB}(u_{ij})$).  For example, for the
 quantum group $SU_q(2)$  one can consider any $(2j+1)$ - dimensional
 representation (see
\eg\ ref [23]). If $j=1$,  this would give us the $O_q(3)$ quantum group \s\
 model.
\hfil\break
b) In the algebraic formulation of the models one has to make precise the
 commutation relations between the quantum group \s\ fields and their
 variations.
 Only when
these are known can we  derive the field equations from the action.\hfil\break
c) The action of the algebraic quantum group \s\ model can be considered as the
argument (after exponentiation) of the generalised Feynman path integral
provided that the suitable  formulae  for the integration over  the quantum
 group
functions $f(G_q)$ are found. This problem bears some analogy with the Berezin
integration rules for Grassmann algebras, and in the case of an arbitrary
 quantum
group, is still to be determined . (See, however, ref [24] for a discussion of
a
 simple
two-dimensional noncommutative case). \hfil\break
d) The concept of a quantum group \s\ field should be useful when
one wants to introduce the notion of generalised gauge fields, with {\bf local}
gauge transformations described by quantum group parameters. In section 3 we
have introduced the $U(1)$ gauge field but because of its Abelian nature
the field's noncommutative aspects were absent. It would be interesting to
consider \eg\ the $U_q(3)$ \s\  model coupled to non-Abelian $U_q(2)$ gauge
fields. This should  allow us (by gauge fixing) to formulate the quantum
group \s\ model on the coset ${U_q(3)\over U_q(2)}$.
\par Finally, we would like to add that although in this paper we have
 considered
\s\ fields taking values in noncommutative algebras the two best known choices,
namely, the quaternionic algebra (see \eg\ ref [25]) and the Grassmann algebra
(see \eg\ ref [26]) are finite dimensional. In the case of quantum groups,
for a generic $q$, the algebra of noncommuting functions $f(G_q)$ is infinite
dimensional and so in order that we can
 extend \eg\ the superfield formalism of
 supersymmetric
theories to the case of quantum groups, the new formal tools still have to be
developed.

\ack
Most of the work reported in this paper was performed when one of us (WJZ)
visited the Weizmann Institute of Science and the other (JL) the University
of Bordeaux I. WJZ wishes to thank the Weizmann Institute of Science
for its hospitality and the Einstein Center
for Theoretical Physics for the support of his stay in Israel.
  JL wishes to thank Prof. Minnaert for the hospitality at Bordeaux I.

 In addition JL would like to thank S.L. Woronowicz for
numerous discussions.

\endpage
\refout
\endpage
\bye